# Study of Feature Importance for Quantum Machine Learning Models

Aaron Baughman,[1] Kavitha Yogaraj,[2] Raja Hebbar,[3]
Sudeep Ghosh,[2] Rukhsan Ul Haq,[2] and Yoshika Chhabra[2]

[1] *IBM, Cary NC, USA*
[2] *IBM Bangalore, KA, India*
[3] *IBM Coppell TX, USA*

Feature or predictor importance is a crucial part of data preprocessing pipelines in classical machine learning. Since classical data is used in many quantum machine learning models, feature importance is equally important for quantum machine learning (QML) models. This work presents the first study of its kind in which feature importance for QML models has been explored and contrasted against their classical machine learning (CML) equivalents. We developed a hybrid quantum-classical architecture where QML models are trained and feature importance values are calculated from classical algorithms on a real-world dataset. This architecture has been implemented on ESPN Fantasy Football data using Qiskit statevector simulators and IBM quantum hardware such as the IBM Mumbai and IBM Montreal systems. Even with the current scale Quantum computers are at, during these experiments, the results are promising. To facilitate current quantum scale, we created a data tiering, model aggregation, and novel validation methods. Notably, the feature importance magnitudes from the quantum models had a much higher variation when contrasted to classical models. We are able to show that equivalent QML and CML models are complementary through diversity measurements. The diversity between QML and CML demonstrates that both approaches can contribute to a solution in different ways. Within this paper we focus on Quantum Support Vector Classifiers (QSVC), Variational Quantum Circuit (VQC), and their classical counterparts. ESPN fantasy football's Trade Assistant with IBM Watson combines advanced statistical analysis with the natural language processing of Watson Discovery to serve up personalized trade recommendations that are fair and proposes a trade. Here, player valuation data of each player has been considered and this work can be extended to calculate the feature importance of other QML models such as Quantum Boltzmann machines.

42## I.  INTRODUCTION

QML is a new paradigm and has shown promise for many domains. QML works fundamentally in a different way than CML, which offers some unsolved challenges as well as advantages. A quantum model encodes the classical data into a multi-dimensional Hilbert space that gives it a rich data representation. The data is represented in quantum space and used by QML models to exploit variational quantum circuits and quantum principles to perform a given task. Today, quantum computing (QC) is limited by noise and hardware capabilities, characterized by the number of qubits, quantum volume and CLOPS speed [1]. The limitations constrain the maximum circuit depth a quantum system can execute, which makes it difficult to process real-world data with a larger number of features.

The current practice in QML is to pass data directly to the QML models after encoding it into quantum states using quantum feature maps. Different than CML and in particular deep learning, QML does apply data pre-processing routines cleansing and extract relevant information from the data. Given the significance of feature importance for QML model performance, we have developed a pipeline for calculating feature importance for quantum kernel-based models and variational quantum circuits models.

In CML, feature selection and feature importance are vital components for good performance of machine learning models. Feature selection helps to extract the relevant predictors while providing a quantitative measure for the importance of each feature in the final model. These feature engineering tasks are equally applicable to QML in which the model is built on a large feature space represented by the Hilbert space spanned by the qubits. QML models interact with classical data through feature maps. As a result, we need to have a data processing and feature engineering pipeline for the QML models. In the current literature, there are proposals to implement quantum principal component analysis (QPCA), quantum SVD, and other similar feature engineering methods [2–4]. However, there has not been an attempt to integrate feature importance algorithms with QML models to explore the role and significance of each feature in the final performance of the QML model. To our knowledge, our work is the first to build a pipeline for calculating the feature importance of QML models. We have integrated permutation importance (PI) and Accumulated Local Effects (ALE) with quantum kernel-based and variational quantum circuit-based models.

We quantify the differences between the feature importance measures created by QML to CML results and subject matter expert (SME) knowledge. We created several diversity measures to study the variations between approaches.

The feature importance created by quantum models created through its rich and diverse Hilbert space should be tested and quantified with respect to classical results and subject matter experts (SME) knowledge. To establish this claim, we created diversity measures to study the variations among all these.

This paper is organized as follows. In section *II* & *III* we introduce classical feature importance and the theory of its applicability. In section *IV*, we described the dataset, quantum models, and implementation techniques to obtain the feature importance. The section *V* describes the techniques to aggregate and normalize data to run quantum models on real datasets, which culminates with the results of our diversity measures in *V I*. Finally, in section *V II*, we summarize our analysis and discuss future work.

## II.  FEATURE IMPORTANCE

Feature importance refers to techniques that assign a score to input features based on how useful they are at predicting a target variable. Feature importance scores play an important role in a predictive modeling projects such as providing insight into the data and model, the basis for dimensionality reduction, and feature independence testing. Feature selection can improve the efficiency and effectiveness of a predictive model. There are many types and sources of feature importance measures such as statistical correlation scores, coefficients calculated as part of linear models, decision trees, and permutation importance scores. A lot of machine learning and in particular, deep learning models, are black-box models. These types of models are not easy to interpret or explain. However, users need to understand why and how a model works. A new field of machine learning, Trustworthy AI, has been developed to increase the transparency, interpretability, and explainability of models [5].

Feature importance is a key technique to understanding the performance of a model. The methods help users understand why, how, and which features are important to a model's performance. Our work extends the current field into QML. However, working with quantum models is a manifold problem that increases the difficulty of obtaining feature importance on real life data [6]. The challenges of working with a quantum model are:

- limited number of available qubits and quantum volume.

- processing real world high dimensional data that grows exponentially on quantum hardware.

# III. METHODS OF QUANTUM FEATURE IMPORTANCE

Throughout this work, we applied two different types of feature importance algorithms:

1. Permutation Importance [7]
2. ALE Accumulated Local Effects (ALE) for feature importance [8]

## A. Algorithm of Permutation Importance

Permutation Importance (PI) is a well-established technique to calculate feature importance in CML [9]. PI works by shuffling predictor values across rows of a particular column to generate a new data set. For example, in Figure 1 the shuffling of a column is shown to generate a modified version of the dataset to calculate the accuracy. 'Accuracy' column represents accuracy score of the dataset and the 'low score' column's rows are permuted and the accuracy of the whole data set is calculated. The permutation feature importance algorithm is based on [11]. This algorithm can be applied with any opaque estimators represented by a fitted predictive model to compute the reference scores of the model on data set [10]. We extend this technique to calculate the feature importance of quantum models with the following steps:

1. For each $feature_j$ of dataset D, where, index the j ∈ [1, total number of features in D]

2. The $n_{repeats}$ is the number of repetitions, where the index k ∈ [1, $n_{repeats}$]. By default $n_{repeats}$ =5 as per [11].

3. For every $feature_j$ of the dataset D, the rows of $feature_j$ are shuffled to generate a modified version of the dataset named $\tilde{D}_{k,j}$ as shown in Figure 1.

4. Compute the scores $s_{k,j}$ for the index k,j for the model $m$ on modified data $\tilde{D}_{k,j}$.

5. Compute importance $i_j$ for feature $feature_j$ defined in equation 1 below:

$$i_j = s - \frac{1}{K}\sum_{k}^{K} s_{k,j}$$

(1)

Figure 1: Permutation Importance applied to ESPN fantasy football data.

## B. Algorithm of Accumulated Local Effects (ALE) for Feature Importance

ALE approach determines how predictors and their correlations influence a model's prediction based on a feature values of local windows. Local windows are created by dividing the interval over which the feature values span, into a grid of smaller intervals. The averaging is done on each one of these smaller intervals which are also local windows [12]. An example of ALE feature importance calculation has been shown in figure 2 in which feature importance is being calculated for X1 feature which is correlated with X2 feature. First of all, the feature X1 is divided into intervals which can be seen as vertical lines in figure. Then, for each data point(feature value) in given interval, the difference in the prediction is calculated bu substituting the upper and lower bounds of that interval(horizontal lines). These differences are then accumulated and transformed to have zero mean. All these accumulated differences give us the ALE feature importance plot. Research groups such as oracle data science have implemented the ALE method [13]. The algorithm was introduced to improve upon the Partial Dependency Plot (PDP) model[14]. Partial dependency plots show the dependence between the target response and a set of input features of interest. PDP rely on marginal distribution and estimates the value of the prediction function [15].

$$f_{s,PDP} = \int f(x_s, X_c) dP(X_C) \qquad (2)$$

This equation gives the value of the prediction function $f_{s,PDP}$ at feature value $x_s$, averaged over all features in $X_C$.

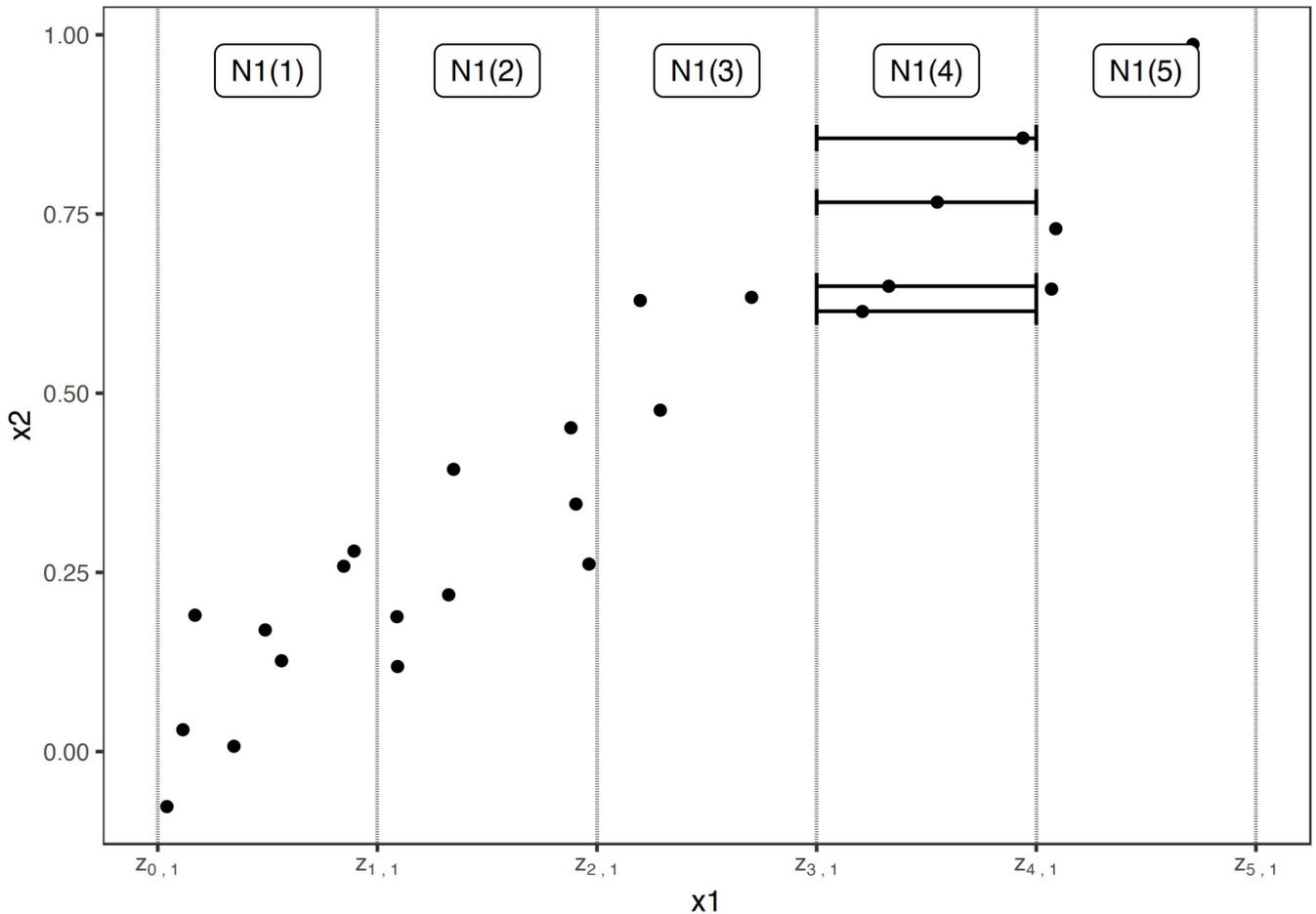

Figure 2: Calculation of ALE feature importance for X1 feature which is correlated with X2 feature

PDPs are easy to understand but cannot be used for correlated features that are generally present in real life data. As a result, ALE was introduced to calculate the feature importance when predictors are correlated. What is common between PDP and ALE methods is that both use marginal distribution over feature space to calculate the average contribution of a given feature to the model prediction. However, in the case of ALE, a local window is chosen to find out the effect on the model prediction.

In ALE, the focus quantity is on change in prediction instead of predictions as used in PDP.

$$\tilde{f}^S_{ALE}(x_s) = \int_{z_{0,s}}^{x_s} \frac{\partial \tilde{f}(x_S, x_C)}{\partial x_S} dP(X_C|X_{S=z_s dz_s}) \tag{3}$$

On quick inspection of equations 2 and 3, one can clearly see that in case of ALE, the averaging is done not on the values of the features rather on the derivatives or changes of the prediction function. The steps of the ALE algorithm are as follows:

1. Select the model which has been used for the predictions.

2. Select the feature for which ALE model will calculate the feature importance.

3. Calculate the upper and lower bounds over the intervals which will be used to calculate the changes in the model predictions.

4. For each interval, sample the conditional distribution. The sampled distribution gives access to the simulated data sets which are used to calculate the differences in model predictions.

5. The prediction differences within each interval are averaged and accumulated in order, such that the ALE of a feature value that lies in a given interval is the sum of the effects of the first through that interval.

6. Finally, the ALE values at each interval are transformed so that the mean effect is zero.

$$f_{j,ALE}(x) = \tilde{f}_{j,ALE}(x) - \frac{1}{n}\sum_{i=1}^{n} \tilde{f}_{j,ALE}(x_j^i) \tag{4}$$

## IV. REAL-WORLD PIPELINE TOWARDS FEATURE IMPORTANCE

This section discusses the input data set, models, and how the feature importance is calculated. We also describe how to work with a large number of features using our feature tiering and aggregation methodology within QML models.

### A. Input Data Set

Entertainment and Sports Programming Network (ESPN) provided Fantasy Football data to support our work. The IBM Trade Assistant interpreted the data to produce an expanded dataset about Fantasy Football trades [16]. The expanded data set contains 146 features that describe the value of a set of players that could be involved in a trade. The valuation about a single player within a trade is extrapolated to derive 146 features about a single football player's valuation relative to all other football players. A total of 4,733 samples were extracted from ESPN-rated trades in the form of labeled data. Trades were given an overall rating between a low score of 1 and a high score of 10. Any trade that was rated 4 or higher was accepted as a good trade. The label was reduced to a binary value that represented either a good trade with a score of 1 or bad trade with a score of 0. The training data was balanced over good and bad trade classes. To overcome the limitations of quantum hardware, we grouped features together based on their correlation to the label. This process is described in section IVC. This divide and conquer approach allowed us to load an overall of 15 tiers into a quantum circuit. A summary of our input data is as follows:

1. ESPN fantasy football trade data which consists of 146 features. The 147 [th] column is the class name (good or bad trade) marked as 1 or 0, correspondingly.

2. Tiering data which maps each feature name to it's corresponding tier. We have overall 15 tiers numbered as 0-14. The first 14 tiers contain 10 features each while the last tier, contains the remaining 6 features.

3. ESPN expert/SME ranked feature list, that was used as a rank-based evaluation metric. This is a list of all the feature names with a corresponding rank assigned to each feature ranging from 1-147.

### B. Selection of backends and Number of Features

In a quantum computing environment, there are many considerations for the execution of the quantum algorithms such as quantum circuit depth, quantum volume [17], and the number of qubits. Quantum circuit depth is the number of time steps or time complexity required for quantum operations within a quantum circuit to run on a quantum device [18]. A Quantum Volume [12] is a metric that measures the capabilities and error rates of the quantum computer that expresses the maximum size of square quantum circuits. The number of qubits is one of the few metrics which helps us to determine how many features can be simulated in a quantum device. At the time of writing this paper couple months after the experiments, the highest qubit IBM Quantum device available is the Eagle processor which has 127 qubits with a quantum volume of 64. This system was not available, in October 2021 while conducting this study and corresponding experimentation. Within our problem, each of the 146 features map to a qubit plus some more qubits to handle errors and processing, that would require a large circuit depth and even more qubits that is not possible today within 127 qubit device. Hence, we would still use tiring methodology to simulate 146 features. As a result, we divide the 146 features into 15 tiers to run on IBM Montreal and IBM Mumbai quantum device.

To start, we pick a logical representation of quantum computing device within Qiskit called a backend [19]. A backend represents either a simulator or a real quantum computer that is responsible for running quantum circuits and returning results. In this paper, we refer to two different backend types:

- Statevector simulator as a quantum simulated backend

- Quantum hardware that uses IBM Mumbai and IBM Montreal with a specification of 27 qubits and 128 quantum volume.

We used the Statevector simulator with 32 qubits to run diagnostic tests of our circuits. The statevector simulator is the most common backend in Qiskit Aer package [19]. The statevector simulator is a tool that runs through the quantum circuit with one shot and returns the qubit statevector with a complex vector of dimensions $2^n$ where n is the number of qubits.

The maximum capability we experimented was 21 features that takes 30 hours to complete on statevector simulator. To process all 146 features, we would consume 250 hours of compute time, which was not feasible. During our experimentation we found optimal tradeoffs for runtime, number of features, and the number of data points. For example, a single tier with 14 features and 4000 data points took 7 hours on statevector simulator backend. Next, we conducted experiments with a 27 qubit, 128 quantum volume IBM Montreal and IBM Mumbai for 14 features and 100 data points which resulted in a quantum depth of 57. We found that PI and ALE algorithm were not able to run on our IBM Quantum devices because the problem space exceeded the quantum depth causing depth recursion errors. However, we successfully experimented with 10 features per tier over 100 data points on both the above mentioned quantum devices. Finally, to summarize the results of our experiments in section VI and maintain consistency, we took 10 feature per tier in both statevector and IBM Quantum device backends. However, the number of data points used in statevector simulator was 4000 whereas 100 data points on IBM Quantum devices.

### C. Feature Tiering

To run on quantum hardware, after the experimentation, 10 features per tier has been selected to cover all the 146 features in the input dataset. In this tiering methodology, we divided the input data into 15 tiers that required 15 different experiments to be conducted. Each of the tiers contains the same number of rows and class labels with 10 features except for the last tier, which has only 6 features. A classical algorithm based on correlation groups the features into tiers. Each feature is ranked based on the correlation it has with it's label. This means that the first 10 features will be more correlated to the label of the data than the last tier. Each of the experiments created a quantum model. Overall, 15 different QML models were created for each type of QML such as QSVC and VQC

Each of the 15 models are ensembled together based on a novel strategy. We developed an approach to aggregate and normalize the output of each model to produce a single result. More details are described in section V and depicted in Figure 3.

### D. Feature Importance Quantum Models:

We are using implicit and explicit approaches of QML techniques to solve feature importance in quantum computing. In the implicit approach, we are using the QSVM algorithm to take advantage of quantum kernel estimation. In the explicit approach we are using VQC. The algorithms can be found in the Qiskit Software Development Kit (SDK)

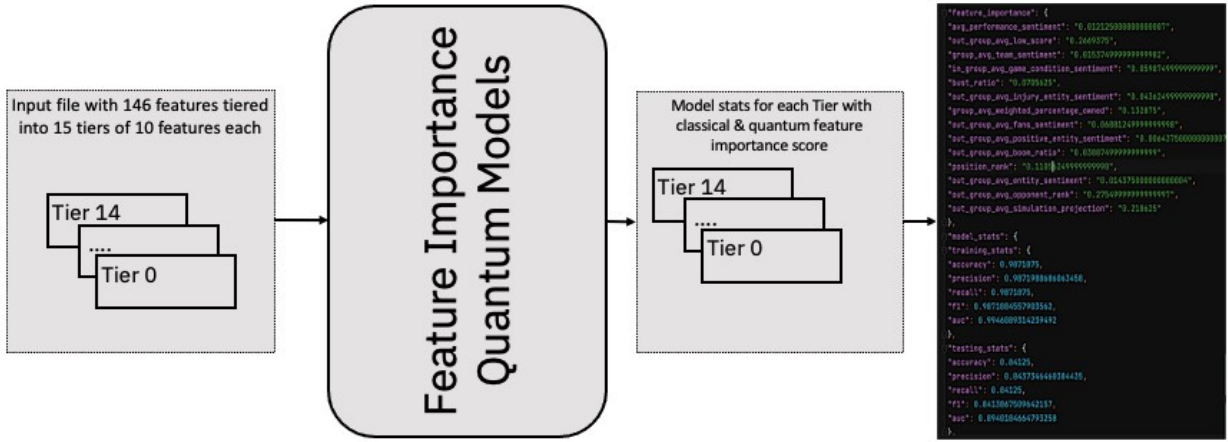

Figure 3: The process of using feature tier groups to establish feature importance.

[19] packages. Figure 4 shows an overview of the quantum approaches we selected. The following sub-sections provide more detail on the QML models.

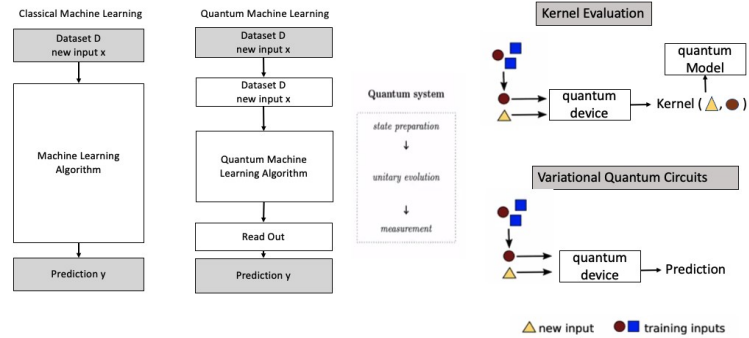

Figure 4: Comparative analysis of CML and QML approaches.

#### 1. Quantum SVM

Supervised classification algorithms are trained on labelled data and are then used to predict labels for new data. This type of training is called supervised learning. Well known classical techniques such as Support Vector Machines (SVM) and Neural Networks (NN) have been trained with supervision and applied to many types of large scale classification problems [21]. In supervised classification, we have a set of points $x$ that are labeled within a particular group. The learning process finds a line or hyperplane that separates the groups. The kernel is a key concept in kernel-based classifiers like SVC. Data cannot typically be separated by a hyperplane in its original space. A common technique used to find such a hyperplane consists of applying a non-linear transformation function to the data. This function is called a feature map. A feature map $\varphi$ is a function that acts as $\varphi : x \rightarrow F$ where $F$ is the feature space. The output of the map on the individual data points, $\varphi(x)$ for all $x \in X$, are called feature vectors. The quantum kernel method can be generalized to the case when the decision function depends non-linearly on the data, by using a kernel trick and introducing a high-dimensional non-linear feature map. The data is mapped from a lower-dimensional space into a higher dimensional non-linear Hilbert-space (H) also known as a feature space. If a suitable feature map has been chosen, it is then possible to apply the SVM classifier to the mapped data in Hilbert space. Classifying data points in this new feature space is observing how close data points are to each other. These collections of inner products between each pair of data points in the new feature space is called the kernel. We implemented feature map circuits, which makes it easier to compute these calculations on a quantum computer rather than doing these computations over a classical computer where it is not efficient. A quantum processor is used to estimate the kernel in the feature space. After the quantum kernel matrix is generated, we need to extract the information about the quantum kernel from the quantum circuit to input it into the classical SVM algorithm. Finally, the SVM uses the non-linear kernel to create a hyperplane which separates the data into classes [6, 20, 21].

We use the ZZFeatureMap with a linear entanglement strategy [20]. As shown in Figure 5, ZZ gates are implemented as a CNOT with a phase gate on the lower qubit followed by a CNOT gate. In Qiskit [19], this circuit is created with 4 features using ZZFeatureMap with the following characterstics: *ZZFeatureMap*(*feature dimension* = 4, *reps* = 3, *entanglement* =' *linear*')

- feature dimension =10, which is the number of features per tier.
- Linear entanglement is implemented with a varying degree of entanglement shown in the circuit.
- we repeat the data encoding step 3 times.

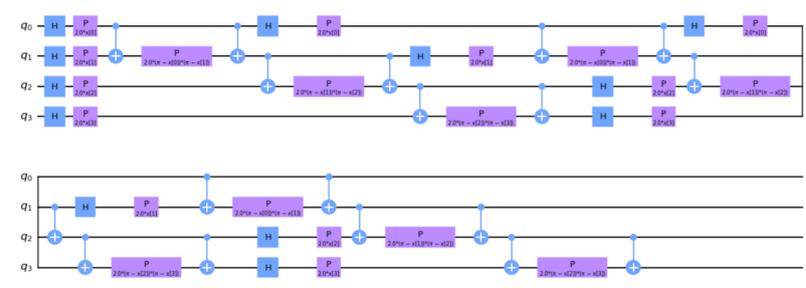

Figure 5: Quantum circuit for ZZfeaturemap with 3 data encoding repetitions over 4 features.

2. *QSVM Applied with Permutation Importance and ALE*

Now that we have defined our QSVM approach, we will apply a feature importance step. Figure 6 depicts the feature importance approach using PI and ALE.

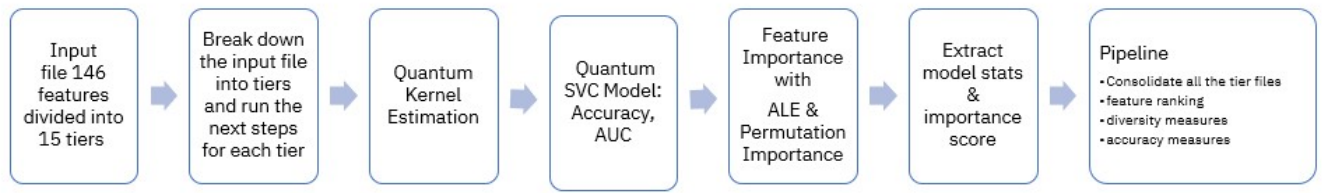

Figure 6: Pipeline for Feature Importance with PI and ALE methods

The following process summarizes our QSVM pipeline.

1. Input exemplars with 146 features that are divided into 15 tiers.

2. Perform data encoding using quantum feature map, ZZFeatureMap, which takes this data and represents it in quantum Hilbert space.

3. With quantum kernel transformer, transform the classical data to a non-linear high dimensional space called the feature space on which a classical SVM creates a hyperplane to separate the labeled samples. The quantum kernel generates a kernel matrix with 80% training and 20% testing from 4000 exemplars.

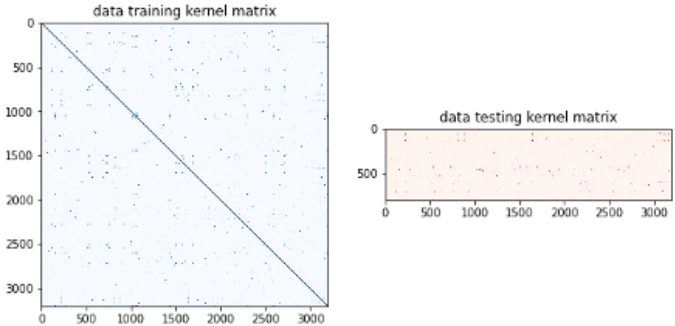

Figure 7: Quantum Kernel Matrix plots for training and test data

4. After calculating the kernel matrix, the pre-computed kernel is used by the SVC algorithm to get prediction accuracy and decision functions.

5. Create feature importance scores for each predictor using PI and ALE. The two approaches are described below.

   PI: Permutation importance is calculated on the quantum model that has been trained. A permutation importance function is used for calculating feature importance, The test kernel matrix will be created for calculation of feature importance using the permutation importance algorithm further explained in section IIIA.

   ALE: From the trained model, we extract the predict function which is then passed on to an ALE Model. ALE model is trained on the training data set from the original model. The feature weights are then calculated using the explained function of the trained ALE model, explained in section IIIB

### 3. Variational Quantum Classifier (VQC)

The second QML model we studied was the VQC algorithm. This approach is based on variational principles which are suitable choices for classification problems within the Quantum Computing era. The three major components of VQC are the feature map, variational circuit, and the classical optimizer. In VQC, classical data is mapped into a higher-dimensional feature space using a feature map where the problem becomes more separable. The layer of the variational circuit is constructed using ansatz gates that provides variational parameters ($\theta$) to perform the learning and tuning of the model. A classical optimizer algorithm is applied to change the attributes of ($\theta$) while minimizing loss. Since quantum computers today are limited by the number of qubits, we cannot work with a large number of qubits like 146 features. Therefore, after tiering the data the resulting dimension of 10 features per tier. To input classical data onto the VQC circuit, we use quantum encoded values using [20]. Figure 5 shows a second-order Pauli-Z evolution circuit that interacts with the classical data to encode it using the feature map connectivity circuit. This circuit has four qubits $|0i^4$ as initial state with the coefficients $\varphi(\sim x) \in \mathbb{R}$, that encodes the classical data $x \in \Omega$ of a Hilbert space. These type of calculations are inefficient and difficult to simulate on a classical computer with bits. Quantum computing provides a computational complexity advantage [20]. We took the following steps to implement a VQC:

- A FeatureMap converted classical data into quantum encodings using the ZZFeatureMap with 10 features, 3 iterations, and a linear entanglement type.

- A VQC is trained and tested. The model returns accuracy scores [14].

- The process flow depicted in Figure 8 was placed into a pipeline with feed forward data.
- A feature importance step such as PI or ALE was applied within the pipeline.

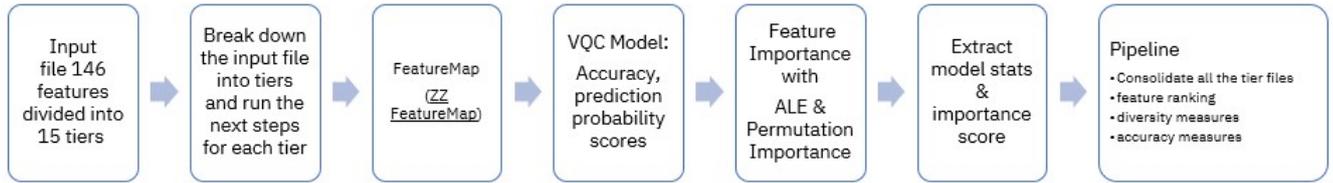

Figure 8: VQC pipeline for feature importance with PI and ALE methods

## V. FEATURE AGGREGATION AND NORMALIZATION

In the previous sections, we discussed how our large feature sets was tiered into multiple groups for the measurement of quantum feature importance. Now that quantum models have generated model statistics and feature importance groups independently, they need to be aggregated together. Feature aggregation is done to aggregate all the individual tier results into a single experimental run and to have one output. We applied a normalization strategy on the aggregated output results which reflects normalization across all the models with one model accuracy. Normalization is done over features importance score to get standardized feature importance scores from different models with different ranges to sum to 1, mentioned in equation 7. The goal of feature aggregation is to have one model statistic, classical feature importance score, and quantum feature importance score. As a result of tiering, aggregating, and normalization steps we achieve feature importance scores from quantum and classical models for all the 146 features relative to each other. Classical method used here is XGBoost Algorithm which can handle 146 feature's feature importance but to give fair comparison tiering method is applied to classical method as well. In the following sections, we describe our algorithm for feature aggregation and normalization.

### A. Feature Aggregation and Normalization Algorithm

Our algorithm manages a set of 146 features, *n*, that are split into *m* tier groups. In the case of quantum, $m \in \{0,15\}$ while in classical $m \in \{0,10\}$. The feature indexes $i, j \in \{1, n\}$ determine the feature member within the feature set. The trained models on a group of features, *m*, is indexed with $k_m$. At the conclusion of our algorithm, each feature, *x*, will have a feature importance from a model denoted by $x_i$.

The tiered feature groups for a quantum model are combined based on model accuracy. Equation 5 shows how each feature importance has been rewarded based on a model's accuracy to produce feature importance reward, (*fir*).

$$fir_i = \frac{1}{2}((e^{accuracy_{k_m}} * x_i) + (\tan(accuracy_{k_m}) * x_i)) + x_i \quad (5)$$

where $fir_i$ = $i^{th}$ feature importance that has been rewarded by accuracy of that model $k_m$ where model km refers to the model at $m^{th}$ tier.

$$fir\_norm_i = \frac{fir_i}{\sum_{j=0}^{n} fir_j} \quad (6)$$

and *fir norm$_i$* = $i^{th}$ feature importance that has been normalized. Now, each feature importance reward needs to be normalized so that the sum of all is equal to 1 as shown in equation 7.

$$\sum_{j=0}^{n} fir\_norm_j = 1 \quad (7)$$

## B. Feature Ranking and Diversity Measures

In this section, we quantify the diversity between classical and quantum importance scores to validate that the explainability of QML and CML is different. ESPN football Subject Matter Experts (SME) provided us with a manual list of feature importance ranks. We contrasted ranks and feature importance magnitudes across classical, quantum, and ESPN SME feature importance values. The differences between the feature importance results indicate that statistical, human, and quantum uncertainty can contribute to an alternative solution to a problem such as generating fantasy football trades between teams. Our new diversity notation depicts 3 measured diversity metrics: accuracy, feature rank difference, and feature magnitude variance for quantum and classical computing.

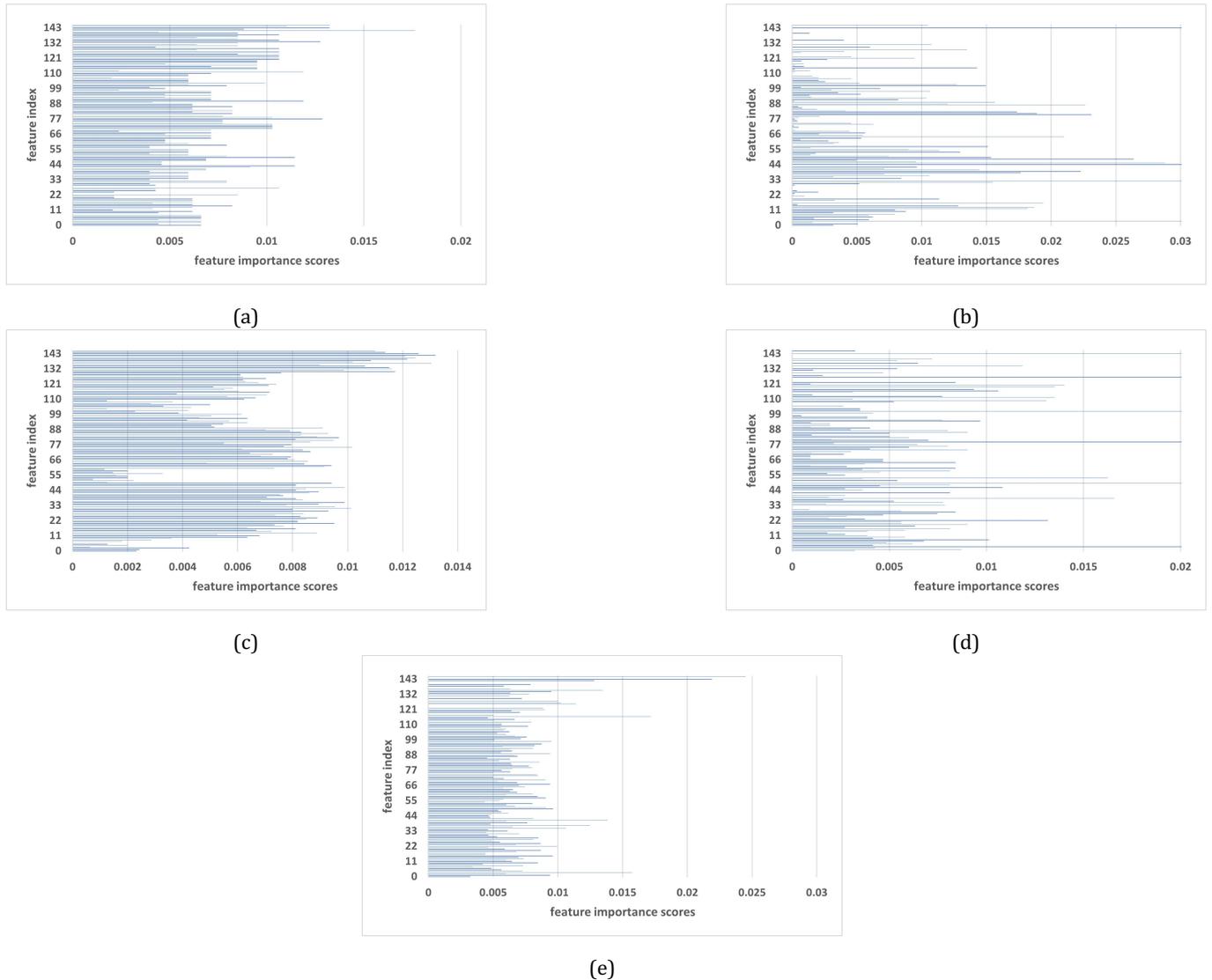

Figure 9: Aggregated and Normalized Feature importance score on a quantum device: (a) Graphical representation of the normalized quantum feature importance score for Model 1: *QSV M + PI* for all features which are referred to by their feature index. (b) Graphical representation of the normalized quantum feature importance score for Model 2: *QSV M + ALE* for all features which are referred to by their feature index. (C) Graphical representation of the normalized quantum feature importance score for Model 3: *V QC + PI* for all features which are referred to by their feature index. (d) Graphical representation of the normalized quantum feature importance score for Model 4: *V QC + ALE* for all features which are referred to by their feature index. (e) Graphical representation of the normalized quantum feature importance score for the baseline classical model: *XG BOOST*.

### 1. Feature Ranking

Feature ranking is the process of ordering the features by the value of feature importance scores returned by PI and ALE methods [10, 13], which usually measures feature relevance. Our experiments provided accuracy measures and normalized feature importance scores from both classical and quantum classification feature importance models. In addition to classical and quantum feature importance scores, we also have the ESPN Subject Matter Expert (SME) importance score as another comparative perspective. The SME feature scores was provided by ESPN experts [16]. As a result, we have three importance scores for each feature. We ranked the individual features based on the quantum, classical, and SME importance score magnitude in descending order. The feature ranks for all the models are contrasted to the feature rank difference between the equivalent classical, quantum, and SME scores. Figure 10 shows the quantum model's ($QSVM + PI$) feature importance which is plotted versus classical and SME rank. The scattering of classical and SME ranks show the differences in feature is ranked across different types of uncertainty. We formalize the notation with the following:

- $q\ rank_i = i^{th}$ feature's quantum rank

- $c\ rank_i = i^{th}$ feature's classical rank

- $sme\ rank_i = i^{th}$ feature's SME rank

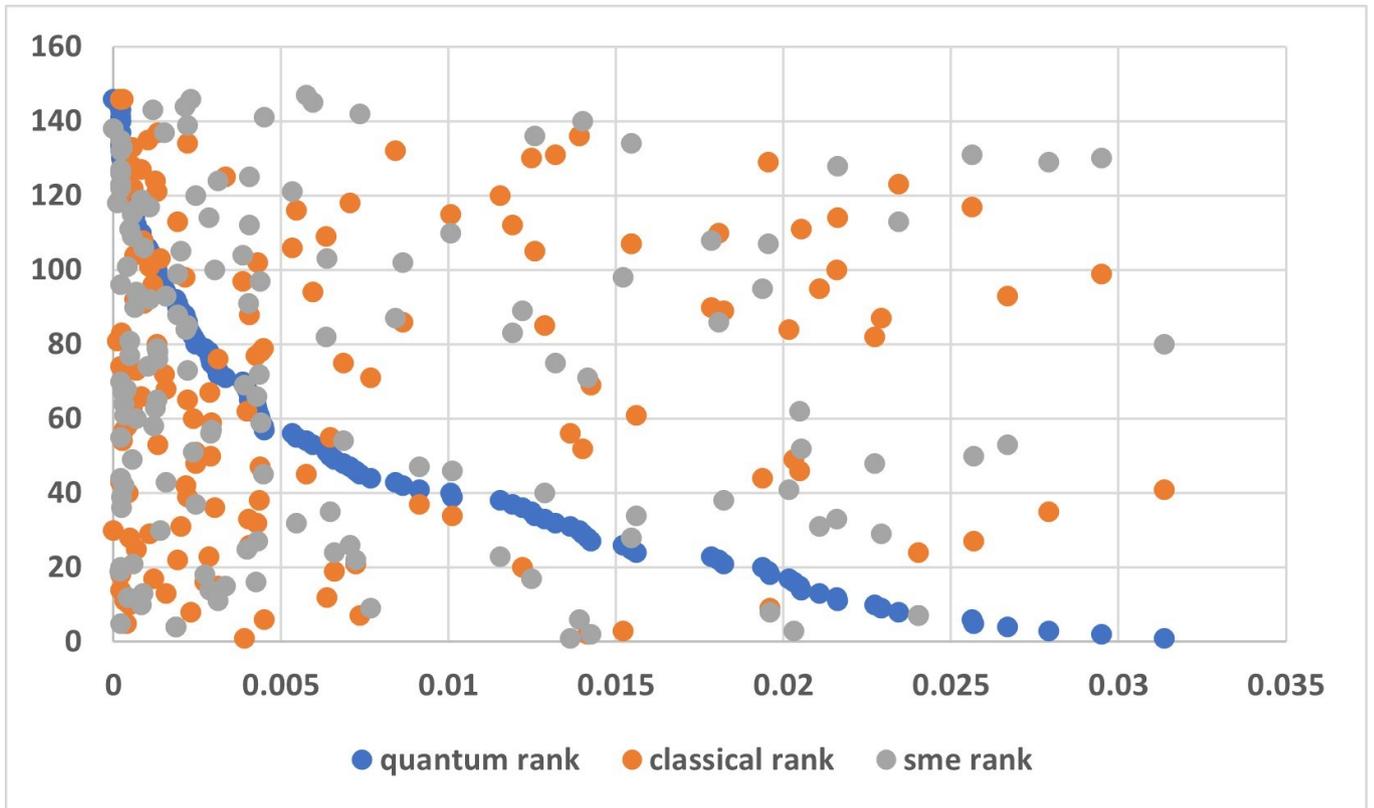

Figure 10: Three different ranks in a single plane where blue represents quantum rank, orange represents classical ranks and gray for SME rank.

### C. Diversity Measures

We measured diversity in terms of accuracy, feature rank difference, and feature magnitude variance to compare similar CML and QML approaches. Measuring the diversity of each individual feature importance score, ranked against its classical and SME counterpart to know how different it is from the rest of the algorithms was important. Over our computing paradigms, we calculate two summary diversity measures: quantum diversity, *qd*, and classical diversity, (*cd*). *qd* is a new diversity measure that has been introduced in this work that not only can be used to compare with classical but also, to trade off feature value importance scores and accuracy. The notation quantum diversity $qd_k$ represents quantum accuracy $qa_k$, corresponding to each model *k*.

$$qd_k = qa_k @ qrank\_diff_{avg,k} @ qvar_k \tag{8}$$

The first term, $qa_k$, is the quantum accuracy for model *k*. Next, *qrank diff*$_{avg}$ is the diversity of feature rank, defined by the average percentage rank difference between the two pairings of quantum with classical, and quantum with SME. The first diversity measure, *qrank diff*$_{avg}$, averages model *k*'s predictor importance percentage rank difference compared against both the SME labeled predictor importance by ESPN, and the classical model feature importance. The quantum rank difference provides a percentage ordinal rank difference measure. To calculate the quantum rank difference, we get the quantum-classical rank difference and quantum SME rank difference as per the equations below:

$$qrank\_diff_{avg} = \frac{qc_{\%rank\_diff} + qsme_{\%rank\_diff}}{2} \tag{9}$$

$$qc_{\%rank\_diff} = \frac{\sum_{i=0}^{n} |q\_rank_i - c\_rank_i|}{max\_rank\_diff} \tag{10}$$

$$qsme_{\%rank\_diff} = \frac{\sum_{i=0}^{n} |q\_rank_i - sme\_rank_i|}{max\_rank\_diff} \tag{11}$$

where,

- *max rank diff*= 10658, a constant value generated by classical feature importance ranking. the equation for calculating *max rank diff* is as below.
- max _rank diff = it calculates the maximal difference if the ranks were completely opposite

$$max\_rank\_diff = \sum_{i=0}^{n} |rank\_asc_i - rank\_desc_i| \tag{12}$$

- *rank asc$_i$* = is the feature importance ranking in ascending order from the classical ranking done by XGBoost • *rank desc$_i$* = is the feature importance ranking in descending order from the classical ranking done by XGBoost
- *i*= the feature index within the feature set.
- *n*= the number of features in the feature set.

The variance of the magnitudes of each predictor provides a second diversity measure called **qvar**. The term **qvar** is the average variance of the feature predictor importance overall quantum models.

$$qvar_k = \frac{1}{n} \sum_{i=0}^{n} (p_i - \mu)^2 \tag{13}$$

where,

- $p_i$= value of the $i^{th}$ predictor importance score.
- $\mu$ = is the population mean.
- *n* = is the number of rows *i*.

14The **cd** evaluation vector shows classical model accuracy, predictor percentage rank difference, and predictor magnitude variance [16].

$$cd_k = ca_k @ \text{crank diff}_{avg,k} @ cvar_k \tag{14}$$

$$crank\_diff_{avg} = \frac{1}{2}(c\_q_{\%rank\_diff}) + \frac{1}{2}(c\_sme_{\%rank\_diff}) \tag{15}$$

$qc_{\%rank\,diff}$ from equation 10, is assigned to $c\,q_{\%rank\,diff}$ as quantum-classical rank difference average is the same as classical-quantum rank difference average.

$$c\,q_{\%rank\,diff} = q\,c_{\%rank\,diff} \tag{16}$$

$$c\_sme_{\%rank\_diff} = \frac{\sum_{i=0}^{n}|c\_rank_i - sme\_rank_i|}{max\_rank\_diff} \tag{17}$$

These are the Constraints and constants for the above calculations:

$$\text{max rank diff} = 10658 \tag{18}$$

$$0 \leq \text{crank diff}_{avg} \leq 1 \tag{19}$$

$$0 \leq \text{qrank diff}_{avg} \leq 1 \tag{20}$$

- quantum accuracy and classical accuracies needs to be between 0 & 1:

$$0 \leq qa_k \leq 1 \tag{21}$$

$$0 \leq ca_k \leq 1 \tag{22}$$

- $1 \leq \text{rank ascending order}_i \leq 146$; where $\text{rank ascending order}_i \in Z$ Ranking of feature importance score is done using ascending order is considered

Equation 23 shows an example the *QSV M +PI* model results for **qd** and the *XG Boost* tree model results for **cd** using our notation. We calculate the rank diff *qrank diff$_{avg}$*, *crank diff$_{avg}$*, and variance *qvar* and *cvar* using the equations 8 to 20. The accuracy of the models are noted as classical accuracy, $ca_k$, of 95.0% and quantum accuracy, $qa_k$, of 83.5%. The rank difference *crank _diff$_{avg}$* is calculated to be 65% where as *qrank diff$_{avg}$* is 63.8%. The feature importance variance for classical, *cvar*, was 0.00262 with quantum feature importance variance, *qvar*, as 0.0648. The results are formally written with the classical diversity and quantum diversity measures as follows.

$$qd = 83.5\%@64\%@0.0648 \tag{23}$$

$$cd = 95.0\%@\,65\%@0.00262 \tag{24}$$

## VI. RESULTS

Quantum feature importance models discussed in this paper have been built using both the statevector *simulator* from Qiskit and also IBM Quantum devices[19]. The output from all the four models: *QSV M +PI*, *QSV M +ALE*, *V QC + PI* and *V QC + ALE*, that includes model accuracy, variance, normalized classical and quantum feature importance scores are used to calculate feature rank and diversity scores explained in section VB. The baseline classical algorithm used for comparison is 'XG BOOST'[22]. Quantum diversity for models in conjunction with its accuracy and variance helped us to assess the performance of each individual model.

### A. Execution Platform and Runtime Details

To build and deploy models, we used docker containers in conjunction with Kubernetes on the Redhat Openshift container platform (OCP) [23–25]. The configuration of the 2 platforms used across experimentation is described with the device specification in Table I. Initial experimentation was performed on the OpenShift Cluster Platform (OCP) but we were also successful in executing the code base on an IBM Cloud instance to test code portability.

| System Configuration Details | | |
|---|---|---|
| Title | OpenShift Cluster Platform (OCP) | IBM Cloud Instance |
| System Information | System: Linux, Release: 4.14.0-115.14.1.el7a.ppc64le, Version: #1 SMP Thu Oct 3 05:32:24 EDT 2019, Machine: ppc64le, Processor: ppc64le | System: Linux, Node Name: quantum-its-instance, Release: 5.4.0-80-generic, Version: #90-Ubuntu SMP Fri Jul 9 22:49:44 UTC 2021, Machine: x86 64, Processor: x86 64 |
| CPU Information | Physical cores: 20, Total cores: 160, Max Frequency: 3624.0, Min Frequency: 3620.0, Current Frequency: 3624.0, Total CPU Usage %: 0.5 | Physical cores: 8, Total cores: 16, Max Frequency: 0.0, Min Frequency: 0.0, Current Frequency: 2394.290, Total CPU Usage %: 3.3 |
| Memory Information | Total: 506.89GB, Available: 474.18GB, Used: 29.59GB, Percentage: 6.5 | Total: 220.22GB, Available: 216.71GB, Used: 1.58GB, Total Mem Usage %: 1.6 |

Tabel I: System configuration details of the execution platform.

The model pipeline is as shown in Figure 6 and 8, steps 3-7 of the corresponding pipeline can be executed both in serial mode and parallel mode. Serial mode passed data from each tier sequentially to the quantum model; while in parallel mode multiple sub-processes were created, each of which would run parallelly on different cores of the execution platform. When executing on Qiskit's statevector simulator, all four models could be executed in both serial and parallel mode [19]. We saw on average a 64% reduction in execution time when using the statevector simulator in parallel execution mode. This speedup is due to parallelized execution on the statevector simulator. Parallel execution could not be demonstrated with IBM Quantum devices because the jobs from each tier can only be executed serially on the selected quantum device. The execution time of each model on IBM Quantum device depended on two factors: wait-time across the system queues due to device sharing and processing time of each model. Also, it is apparent from experimentation that for practical problems, it is faster to implement this on a simulator as opposed to running in real quantum devices. The execution time specifics are graphically represented in Figure 11.

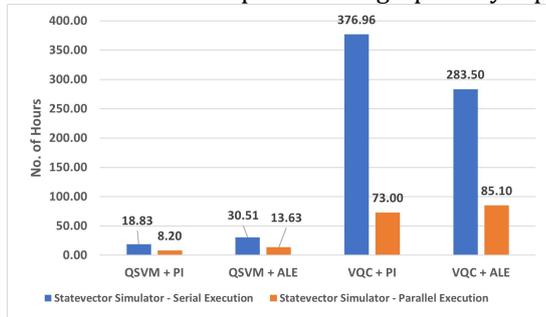
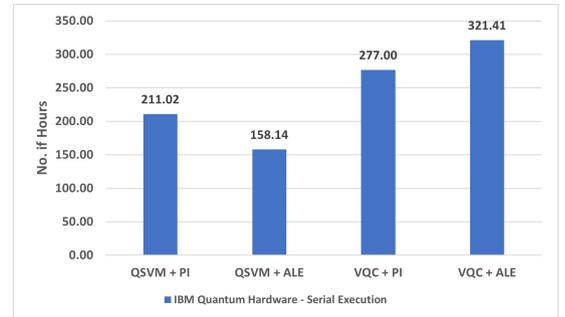

(a)  (b)

Figure 11: Execution run-time of all models - (a) Graphical representation of the execution time of all four models on Qiskits statevector simulator, executed in both serial mode and parallel mode[19]. (b) Graphical representation of the execution time of all four models on IBM Quantum device IBM Montreal and IBM Mumbai., executed in serial mode.

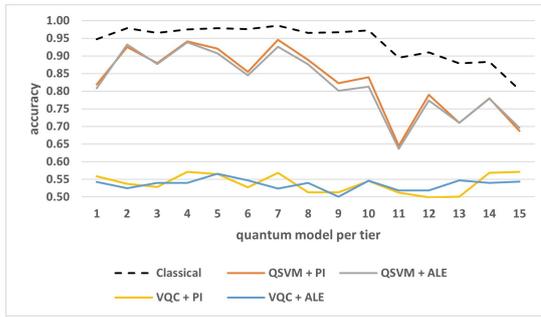
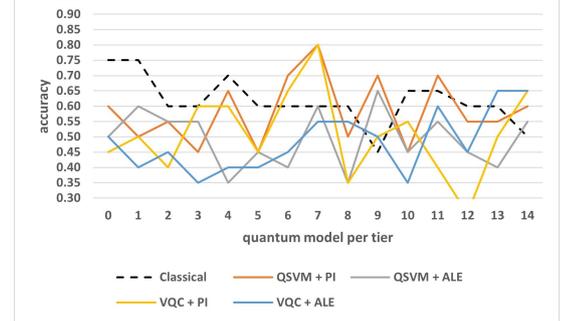

(a)  (b)

| Model | Accuracy | F1 Score | Recall | Precision |
|---|---|---|---|---|
| QSVM + PI | 0.8347 | 0.8307 | 0.8301 | 0.8313 |
| QSVM+ ALE | 0.8256 | 0.8223 | 0.8214 | 0.8232 |
| VQC + PI | 0.5381 | 0.5395 | 0.5388 | 0.5403 |
| VQC + ALE | 0.5359 | 0.5385 | 0.5361 | 0.5409 |

| Model | Accuracy | F1 Score | Recall | Precision |
|---|---|---|---|---|
| QSVM + PI | 0.5819 | 0.5586 | 0.5833 | 0.5359 |
| QSVM+ ALE | 0.4921 | 0.4497 | 0.4933 | 0.4132 |
| VQC + PI | 0.5068 | 0.5254 | 0.5100 | 0.5418 |
| VQC + ALE | 0.4778 | 0.4869 | 0.4833 | 0.4906 |

(c)  (d)

Figure 12: Accuracy plots and statistics of each model. (a) and (c) refers to results from statevector simulator for all four models, dotted black line represents classical XGBoost accuracy plot and the remaining is represented by their respective model names. (c) represents evaluation metrics of model simulation is represented by accuracy, F1 Score, precision, and recall. (b) and (d) contains accuracy plots and evaluation metrics of model simulation on IBM Quantum devices IBM Montreal and IBM Mumbai.

### B. Model Accuracy on Statevector Simulator and IBM Quantum device

The most commonly used metric to evaluate a model is accuracy. However, in our case we wanted diverse results and let a downstream system filter out low quality fantasy football trades. Figures 12 show the accuracy results of our 4 models with balanced classes on 4000 exemplars with 10 features per tier. The train set was 80% of all exemplars. Breaking the results down more, figures 13a and 13c show the model metrics on the the Qiskit statevector simulator [19]. The training and testing accuracy of each tier was plotted in comparison to the baseline classical algorithm 'XG BOOST'. As can be seen on the statevector simulator, the $QSVM + PI$ has the highest accuracy. Both $QSVM +PI$ and $QSVM +ALE$ have a similar accuracy, as the underline QSVM model is the same. The VQC models with PI and ALE had nearly identical results.

To extend our experimentation and validity of our results, all four models were trained and tested on real quantum devices using IBM Mumbai and IBM Montreal. Both of which are IBM Quantum devices with 27 qubits and 128 quantum volume. Permutation Importance algorithm has been observed to consume higher quantum volume but the quantum depth of the circuits was 57. The accuracy plots and measures on IBM Quantum devices are depicted in figures 12b and 12d. Due to the device capacity of the available QPU and our limited access time to the systems shown in Figure 11b, we had to limit our experimentation on QPUs to 100 exemplars.

Our large implementation of a production-grade code base and models on a quantum device was a big accomplishment within the field of quantum computing. We believe the accuracy statistics will continue to increase with additional training data and training epochs.

### C. Ranking and Diversity Model Evaluation

The feature importance output from each model pipeline was used to calculate the feature rank and variance diversity measures. The variability in the feature importance on IBM Quantum device is in line with our experiments on the *statevectorsimulator*, which demonstrates the diversity of our results. What is apparent is that each model proposes a sense of diversity to the feature importance by examining the feature ranking calculation as described in section V. The plots in Figure 13a and 13b are a visual representation of the diversity across feature importance from all four models. Figure 13c and 13d provide tabular information on the quantum diversity measures we discussed in section VC. These measurements provide an extremely valuable insights on how to best evaluate each model.

Assessing inputs from diversity measures of the models on Qiskit's *statevectorsimulator* as mentioned in figure 13c provide us further insight that the *QSV M + PI* model is more ideal than the *QSV M + ALE* model [19]. Not

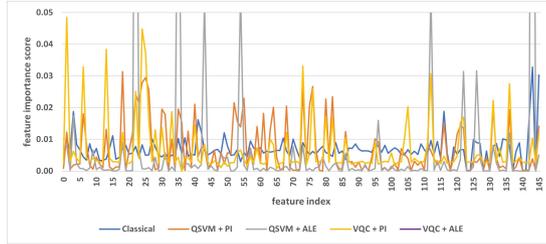
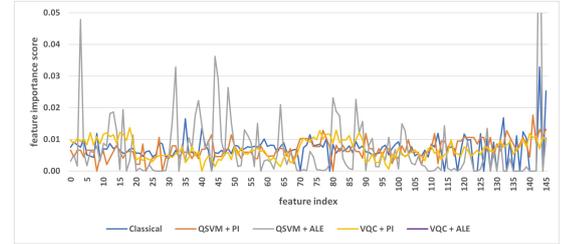

(a)                                                                                (b)

| Model | $qa_k$ | $qrank\_diff_{avg}$ | qvar |
|---|---|---|---|
| QSVM + PI | 83.5% | 63.8% | 0.0648 |
| QSVM + ALE | 82.6% | 57.9% | 0.0214 |
| VQC + PI | 53.8% | 63.1% | 0.0436 |
| VQC + ALE | 53.6% | 58.2% | 0.0070 |

| Model | $qa_k$ | $qrank\_diff_{avg}$ | qvar |
|---|---|---|---|
| QSVM + PI | 58.2% | 64.8% | 0.0260 |
| QSVM + ALE | 49.2% | 66.1% | 0.0131 |
| VQC + PI | 50.7% | 66.4% | 0.0481 |
| VQC + ALE | 47.8% | 59.7% | 0.0111 |

(c)                                                                                (d)

Figure 13: Diversity output from all four models: (a) and (c) represents the diversity across all four models outputs on Qiskit's[19] *statevectorsimulator*, while (b) and (d) represent the diversity across IBM Quantum devices. (a) is a graphical representation of normalized feature importance score of all four models where x-axis represents the feature index and y-axis represents the feature importance score of the respective model on Qiskit's[19] *statevectorsimulator*. (c) is a tabular representation of the output from quantum diversity calculation Qiskit's[19] *statevectorsimulator*. (b) is a graphical representation of normalized feature importance score of all four models where x-axis represents the feature index and y-axis represents the feature importance score of the respective model on IBM Quantum device. (d) is a tabular representation of the output from quantum diversity calculation on IBM Quantum device.

only does it have higher accuracy value, $qa_k$ of 83.5%, the *QSV M + PI* model also has a higher *qvar* of 0.0648, as mentioned in figure 13c. Similarly, in the case of IBM Quantum device experimentation, diversity measures shown in figure 13d provide us further insight that the *QSV M + PI* model is most ideal due to highest accuracy value. It's important to note that higher variance, *qvar* becomes the determinant in case of similar accuracy values.

To give an ordinal comparison between feature ranking done by classical and quantum computing, the top 10 features from all models are depicted in figures 14 and 15. When it comes to feature importance, higher feature importance implies that the feature are more weighted and adds value to the model. It is evident that both the "number of relevant features" and "most relevant features" are varying by model. The varying "number of relevant features" are reflected in figures 13a and 13b, while varying "most relevant features" are reflected in figures 14 and 15. The ranked feature importance output from each model was used to measure quantum diversity. Each model proposes a sense of diversity to the feature importance output, which provides an immense value to the overall objective of feature importance calculation.

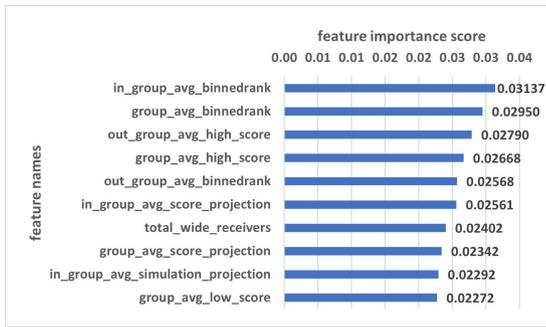

(a)

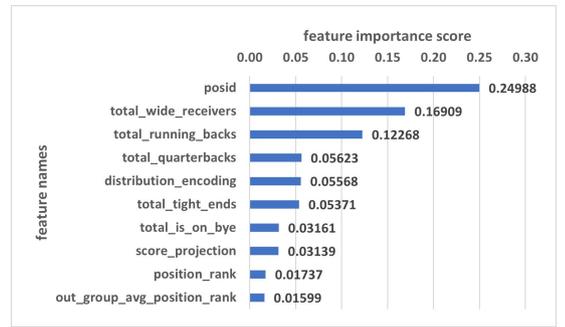

(b)

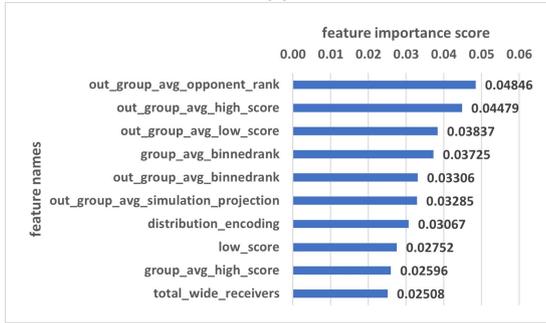

(c)

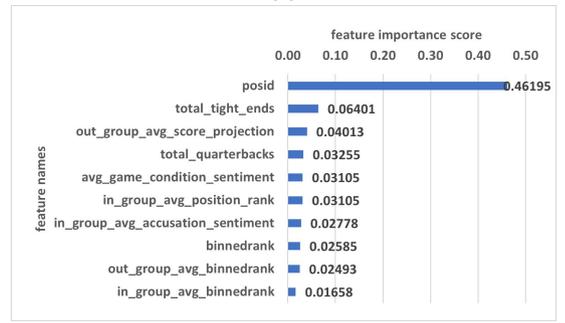

(d)

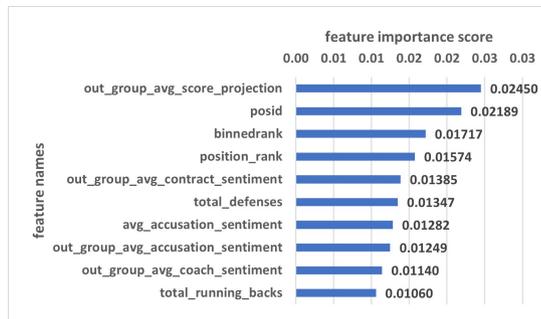

(e)

Figure 14: Graphical representations of the top 10 features after feature ranking on *statevectorsimulator*. The x-axis represents the top 10 features after ranking and y-axis represents the corresponding feature importance score.
(a) contains data from *QSV M + PI*, (b) contains data from *QSV M + ALE*, (c) contains data from *V QC + PI*, (d) contains data from *V QC + ALE*, (e) contains data from the baseline classical model : *XG BOOST*.

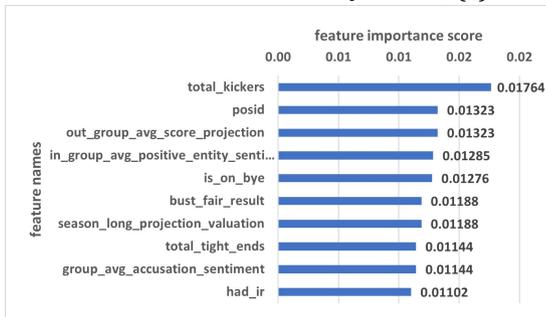

(a)

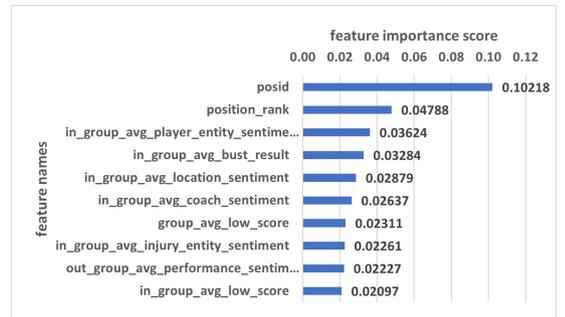

(b)

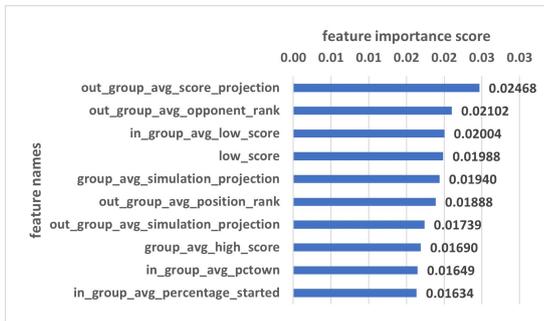

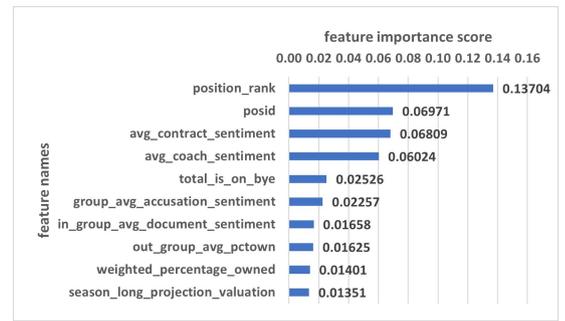

(c)

(d)

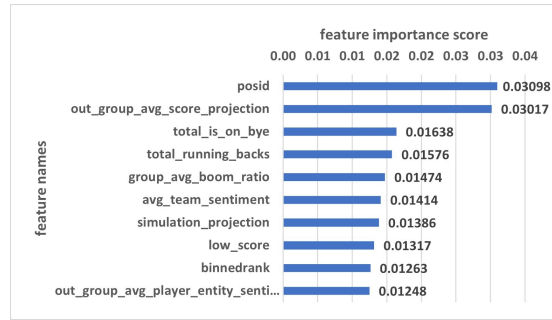

(e)

Figure 15: Graphical representation of the top 10 features after feature ranking on IBM Quantum devices. The x-axis represents the top 10 features after ranking and y-axis represents the corresponding feature importance score. (a) contains data from *QSV M + PI*, (b) contains data from *QSV M + ALE*, (c) contains data from *V QC + PI*, (d) contains data from *V QC + ALE*, (e) contains data from the baseline classical model : *XG BOOST*

## VII.   CONCLUSION & FUTURE WORK

Quantum machine learning has become one of the very important areas for the applications of quantum computing era. Many different kinds of quantum machine learning models has been developed with good successes. However, for the practical and real life applications of these QML models, we also need methods and algorithms which can do the necessary data transformations to enhance the performance of the QML models. One such data transformation is calculating feature importance which helps to identify the relevant and most important features in the given data set for the QML model. In this work, we developed methods for calculating feature importance for quantum machine learning models. Such approaches did not exist in research literature to the best of our knowledge. Our work opens up a new research direction for developing and integrating unsupervised machine learning and data prepossessing algorithms with quantum machine learning models which can enhance the performance of the QML models by many folds.

In this work, We have integrated the feature importance algorithms such as PI and ALE with QML models. Our work has analyzed the practical methods of determining feature importance from quantum kernel and variational quantum circuit models. We have developed a novel pipeline that can be used on Quantum Computing devices. Our pipeline can be extended to other QML models such as QNN, Quantum Boltzmann machines, and etc.

The implementation of our entire pipelines are broadly applicable to Qiskit's [19] statevector simulators and IBM Quantum devices IBM Mumbai and IBM Montreal. The results obtained from our experiments are very interesting as the feature importance's from these models are very different from classical ML models. This gives rise to the diversity in the feature importance and other diversity measures as was desired for the ESPN data.

This indicates that the rich Hilbert space representations of data create a very different representation of feature importance from the models. It was very interesting to find that quantum and classical models are complementary and can produce very different yet meaningful results [16]. In future, we would like to extend our feature importance pipeline to other QML models like as Quantum Boltzmann machines, quantum Generative Adversarial Networks(QGANs). We would also like to test our pipeline on quantum hardware with higher quantum volume so that we can work with larger data-sets.

# Acknowledgement

We would like to thank ESPN, Michael Greenburg, Matthew Berry, Field Yates, Stephania Bell, Daniel Dopp, and Eli Manning for promoting and supporting our work. In addition, our gratitude goes Arjun Kashyap, Daniel Fry, Nicolas Robles, Noelle Ibrahim, Bruce D'Amora, Heather Higgins, Noah Syken, Elizabeth O'Brien, John Kent, Tyler Sidell, Stephen Hammer, Micah Forster, Eduardo Morales, and the IBM Quantum team.